\documentstyle[aps,multicol,prb,epsfig]{revtex}
\begin{document}

\title{Phase diagram of dilute polyelectrolytes: Collapse and redissolution
by association of counterions and coions}

\author{ Francisco J. Solis \\
{\it Department of Physics and Astronomy, Arizona State
University,  P.O. Box 87-1504, Tempe, AZ 85287. \\
Francisco.Solis@asu.edu}}
 \maketitle

\begin{abstract}{Dilute solutions of strongly charged polymer
electrolytes undergo, upon addition of multivaltent salt to the
solutions, a phase transition from extended conformations to
collapsed or bundled ones. Upon further addition of salt they
experience a second transition, a redissolution back into extended
conformations. This paper presents a theoretical study of the
structure and properties of the phase diagram of these solutions.
On the basis of simple phenomenological observations a schematic
phase diagram is constructed that allows a simple and explicit
determination of the direction of the tie lines within the
coexistence region. The actual shape of the coexistence boundary
is determined by means of a model mean free energy functional that
explicitly includes the possibility of association of both
counterions and coions to the electrolyte. It is found that it is
possible to redissolve the electrolytes into conformations where
the bare charge of the electrolyte is overcompensated by the
counterions but, due to the associated coions, can have either
sign of total effective charge. When coion association is
possible, the redissolution approximately coincides with the
reassociation of the coions and counterions in the bulk of the
solution.}
\end{abstract}

\begin{multicols}{2}

\section{Introduction}

Dilute solutions of strongly charged polymer electrolytes
(polyelectrolytes) such as single and double stranded DNA
\cite{widom,rau,bloom,raspaud,raspaud2}, polystyrene sulphonate
\cite{polyesti}, and other biological and synthetic polymeric
systems \cite{raspaud2,rapaudprevious} exhibit phase and/or
structural transitions upon addition of multivalent salts to the
solutions. First, the addition of multivalent salt causes a
collapse of the extended structure of flexible polyelectrolytes or
the bundling in mono- or multi- molecular structures of
semiflexible and rigid electrolytes that segregate into a dense
phase, a phenomenon which in the case of DNA is also know as DNA
condensation. Upon further addition of salt, the electrolytes
redissolve acquiring again extended or unbundled states.  There
are many different examples of this type of system and it is
clear that while the strong charge of the electrolytes is
required for the transitions, the precise geometry of the
polymers is not a crucial factor. The fact that the transitions
are driven only by a combination of electrostatic interactions
with entropic and steric effects has been shown by many computer
simulations. \cite{linse,stevens,thiru}

The first transition to a collapsed state is explained by the
onset of attractive interactions between  almost neutral
structures formed by electrolytes whose charge is almost
compensated by associated multivalent counterions.
\cite{shklprl,solis1,rouzina}
 The redissolution transition indicates that as
the concentration of multivalent salt increases the system finds
more energetically favorable conformations without packaging its
structure. The model presented in this paper emerges from two
previously proposed ideas regarding the redissolution transition.
In a previous work \cite{solis2} it was argued that a determining
factor for the redissolution is the nature of the salty
environment into which the electrolyte redissolves. In particular,
it was emphasized that at large multivalent salt concentrations
the thermodynamics of the multivalent counterions and monovalent
coions is affected by the Bjerrum association \cite{bjerrum} where
the ions of the salt within the solution become strongly
correlated with each other. Thus, an scenario for the
redissolution is that the multivalent ions associated with the
electrolyte prefer, without leaving the polymer, to be exposed to
such  environment. On the other hand, it has been argued by the
group of Shklovskii \cite{rouzina} that the main reason for the
redissolution is the decrease in energy available to the system by
overcharging the electrolytes \cite{survey}.  Since the collapsed
structure has to be effectively neutral, the electrolytes
redissolve so that more counter-ions can associate to them. Both
these ideas can be combined and the result provides an interesting
picture of different possible charged states for the redissolved
electrolyte.

The central point of this paper to exhibit, by means of a simple
model, that for large amounts of multivalent and monovalent salt
added to the system, the  redissolution transition is marked by
the onset of reassociation of the multivalent salt ions on the
bulk of the solvent. In its redissolved state the electrolyte
acquires a large number of multivalent counter-ions that
overcompensate its bare charge but, also, associated with them, a
large number of coions that form a second associated layer. Due
to the large amount of screening of the electrostatic interaction
near redissolution conditions, the decrease in energy due to the
reassociation of coions with the multivalent counterions at the
electrolyte surface is of magnitude similar (but smaller) to the
energy of reassociation in the bulk of the solution. Thus, this
second layer of associated coions can only exist if Bjerrum
association is already present in the bulk of the solution. This
second layer enhances the decrease in energy of the electrolyte
due to the overcharging of the first layer. The presence of the
coions reduces the net charge of the region near the electrolyte
and thus decreases the Coulombic barrier to multivalent ions
overcharge. Furthermore, in such a state, there is a greater
number of particles in strongly correlated states thus decreasing
the energy per electrolyte and making its redissolution more
favorable. The total charge of the near region of the electrolyte
that contains all these associated charges can have either charge
sign, but in all cases the charge from the multivalent ions
overcompensates the bare charge of the electrolyte. This
multiple-species association has been observed in recent computer
simulations by Tanaka and Grosberg \cite{tanaka} where the
electrolyte is a small colloidal sphere and by Holm and Kremer
\cite{holm} where the electrolyte is a rigid rod \cite{holm}. The
possibility of this second associated layer has been known for
long time \cite{lozada,kell} but its implications for these phase
transitions has not been fully explored. Figure 1 presents an
sketch of different possible charged states of the near region of
an electrolyte. Consider, for example, a section of electrolyte
with total bare charge $-12$, and multivalent counter-ions of
charge $z=+3$. At low amounts of multivalent-salt the total
charge of the system is negative, but partially compensated by
(a) positive monovalent or (b) multivalent ions. The collapsed
state (c) is neutral with the electrolyte charge compensated by
positive multivalent ions, where different electrolytes or
sections of electrolytes share the positive charges. In the
redissolved state the charge of the electrolyte is
overcompensated by multivalent ions but the total associated
charge can be (d) positive, (e) neutral, or (f) negative due to
the association of a second layer of negative ions.

\begin{figure}
\psfig{file=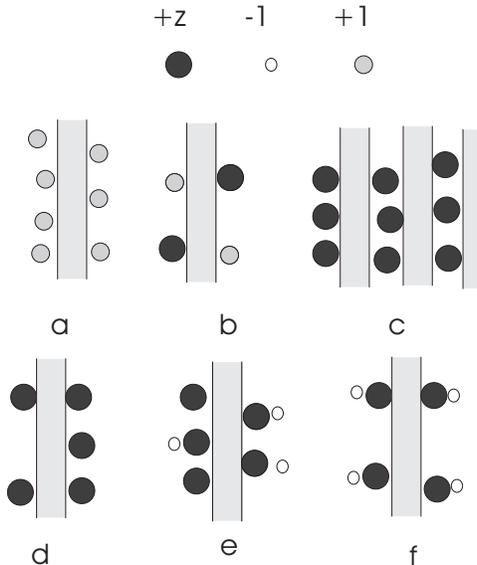,width=2.5in,angle=0} \caption{Sketch of
different charge states of a rod-like electrolyte with associated
counterions and coions. At low amounts of multivalent-salt the
total charge of the system is negative, but partially compensated
by (a) positive monovalent or (b) multivalent ions. The collapsed
state(c) is neutral with the electrolyte charge compensated by
positive multivalent ions. In the redissolved  state, the total
charge can be (d) positive, (e) neutral, or (f) negative.}
\end{figure}

 A second goal of this article is to present a simple
construction of the phase diagram of the multi-species electrolyte
system. While conditions for collapse or redissolution implicitly
assume some form of phase transition its global structure has not
been previously considered in detail. This article proposes, on
the basis of simple phenomenological observations, a construction
of the structure of the phase diagram, the location of the
coexisting phases, and the slopes of tie-lines.

In the rest of the paper, all free energies are measured in units
of $kT$, with $k$ the Boltzmann constant and $T$ the temperature.
All concentrations are presented as number densities. Lengths are
measured with respect to a microscopic length unit $a$ that is the
average size of monomers of the polymer and ions, which are
assumed to be of roughly the same size. The strength of the
electrostatic interaction is quantified by the Bjerrum number $B=
e^2/\varepsilon a kT$ which is energy of interaction of two
elementary charges at its nearest possible distance in water (or
the medium considered).

\section{Structure of the Phase Diagram.}

\subsection{Boundaries of the phase diagram}

The basic components of the system are: a solvent, a dilute
polyelectrolyte with monomer concentration $\phi$, monovalent salt
$s$ and a multivalent salt $\rho$. Since each of these components
dissociate into different ions, we actually start with six
different ionic components in solution. To reduce this number, it
is possible to consider the case in which the charges of the ions
are $-1$ for each monomer of the polymer, $+1$ for the original
counterions of the polymer, $-1$ and $+1$ for the ions of the
monovalent salt, and $+z$ and $-1$ for the ions from the
multivalent salt. It is further assumed that all of the $+1$ ions
are identical, and similarly for the $-1$ ions. Thus there are
only four components, besides the solvent: dissociated polymer at
concentration $p=\phi$, positive multivalent ions $m=\rho$,
positive monovalent ions $t=\phi+s$, and negative monovalent ions
$u=zm+s$. A further constraint is that each phase formed has to be
electroneutral so that the condition $-p+zm+t-u=0$ is always
satisfied.  These restrictions allow for a description of all the
ionic species in terms of only three concentrations, either of
salts $(\phi,\rho,s)$, or ions ($p$,$m$,$t$).

The allowed values of the concentrations of ions can be described
as follows. Each ionic particle density should be non-negative,
and in the reduced representation $(p,m,t)$ the requirement $u\geq
0$ becomes the constraint $t-p+zm\geq 0$. Carrying this constraint
to the salt representation shows that $(\phi,\rho,s)$ should
satisfy

\begin{eqnarray}
0 &\leq& \phi, \\ %
0 &\leq& \rho, \\ %
-\min(z\rho,\phi) &\leq& s. %
\end{eqnarray}

This last inequality has the following interpretation. Consider
the case of a state with a small amount of solvent and a dense
mixture of only polymer ions with multivalent ions, so that
$(p,m,t)=(x,-x/z,0)$ (and $u=0$). This state is physically
realizable but its description in terms of the salt variables is
$(\phi,\rho,s)=(x,x/z,-x)$. To obtain this state one can start
with a mixture of polyelectrolyte and multivalent salt and then
extract monovalent salt, that is, extract both positive monovalent
counterions (given by the polyelectrolyte), and negative
monovalent ions (given by the salt). The boundaries of the
possible states of the system in the salt variables representation
are sketched in Figure 2.

\begin{figure}
\psfig{file=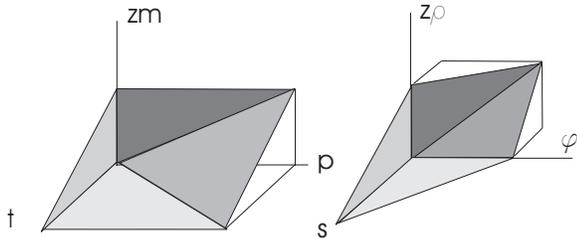,width=3.0in,angle=0} \caption{These diagrams
show sections of the planes that bound the region of allowed
states for the system in the $(p,m,t)$, and $(\phi, \rho, s )$
representations. As explained in the text, the allowed states
extend into negative values of the monovalent salt variable $s$.}
\end{figure}

\subsection{Location of the dense phase} Experiments have shown
the phase segregation of neutral complexes with little amount of
solvent from the bulk of the dilute polyelectrolyte solution. The
precise location of this dense phase in the salt or ionic
components diagram has not been precisely determined
experimentally. Here it is only necessary for our purposes to
assume that this phase has zero or very small amount of monovalent
ions present and that is rich  in polyelectrolyte neutralized with
multivalent ions with a very small amount of solvent. The dense
phase is not a unique point in the phase diagram, but it can be
described as a region in the neighborhood of the point%
$(p,m,t)=(\Phi,\Phi/z,0)$,%
or %
$(\phi, \rho, s)=(\Phi,\Phi/z,-\Phi)$,%
where $\Phi$, is the (large) concentration of monomers in this
phase and, as explained above, the negative value of $s$ reflects
the depletion of monovalent ions from these states. To simplify
the model, however, it can be assumed that the dense phase
consists of only one point, namely $(\phi, \rho,
s)=(\Phi,\Phi/z,-\Phi)$.

\subsection{The tie lines} The precipitation of the collapsed
neutral structures from the dilute solution indicates that the
system has entered into a coexisting phase. A point
$(\phi,\rho,s)$ inside the coexistence region decomposes into two
coexisting points with compositions $(\Phi,\Phi/z,-\Phi)$, namely
the dense state, and a second polymer dilute point of the form
$(\phi-x\Phi,\rho-x\Phi/z, s+x\Phi)$. The assumption that most of
the high polymer density points of the region of coexistence lie
in a very small region and are modeled as one single point
immediately gives rise to this specific decomposition, and it is
clear that the slope of the tie-lines is always determined by this
construction.

To summarize, the phenomenological observation that the dense
polymer phase is neutral with the polyelectrolyte charge
neutralized by multivalent salts leads to a phase diagram of the
form shown in Figure 3 including the tie lines shown. (The choice
of cross section for different values of the $t$ variable are
meant to aid visualization of the diagram and do not reflect
actual energetic considerations. An example of actual
cross-sections in logarithmic variables is presented in the Figure
4).

\begin{figure}
\psfig{file=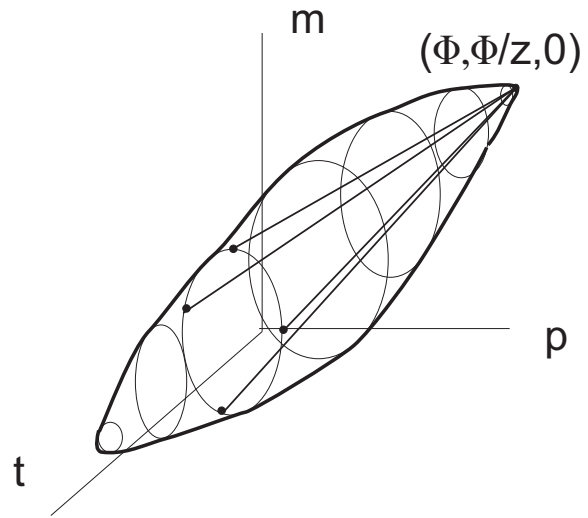,width=3.0in,angle=0} \caption{Scheme of the
region of coexistence in the reduced description of the components
of the system. It is assumed that all tie lines connect points of
the bounding surface of the coexistence region with a single point
$(\Phi,\Phi/z,0)$ at the one extreme of the coexistence region.
Four tie lines are shown for the points corresponding to a section
defined by a constant value of the $t$ component.}
\end{figure}

\section{Free Energy Functional}

To be able to write down an effective free energy functional that
describes the polyelectrolyte phase transition it is necessary to
consider coherently but separately the contributions to the free
energy from three different regions. There is a dense precipitated
phase that is relatively poor in solvent and rich in polymer and
multivalent ions. Within the dilute phase, there are two distinct
regions: the one occupied by the polymer and its associated
charges, and the region away from the polymer occupied only by
solvent and ions.

There are many variants of the problem all of which  present
similar behavior. The same type of phase diagram occurs for
flexible and rigid polyelectrolytes of different lengths. It is
simplest to write a model for a collection of rigid rods (say
short or medium size DNA). In this case the main barrier arising
from the structure of the polyelectrolyte to the formation of a
collapsed state is purely entropic. If one considers very long
semiflexible molecules, the collapse is monomolecular and the
barrier is of a elastic nature instead of entropic, and requires
the introduction of further parameters. The results are
nevertheless similar.

\subsection{Two States modeling}

While it is clear that the region of phase coexistence does not
separate two topologically disconnected regions in the full phase
diagram, it is nevertheless useful to write down two separate free
energy functionals for the states at the ends of a tie line. It is
very hard to write down one single free energy functional that
interpolates between the dense and dilute phase, especially for
flexible electrolytes. When using different functionals for each
phase, a phase transition can be properly determined only if both
functionals are good approximations of the absolute free energy
and refer to a same (arbitrary) zero energy point. In this case
this goal is achieved by using similar approximations in the
calculation of the free energy in the two states.

\subsection{The Dense Phase}

In the dense phase, the charges are very close to each other
although it is likely that they retain a large amount of water.
This phase is electroneutral and effectively forms a bulk region
where the all the charges are strongly correlated with their near
neighbors. In previous work \cite{solis1} it was argued that a
suitable free energy per unit volume $F_{d}$ for this state, is
given by:
\begin{equation} F_{d}(\Phi, \Phi/z, -\Phi)= -M\Phi.\end{equation} %
where $M$ in an effective Madelung constant that depends on the
specific geometry of the polymer and ions used but that reflects
the strongly correlated state of the charges in that phase.

\subsection{The dilute region}
The dilute phase is not homogeneous at a microscopic level. The
charge distribution is very different near the polyelectrolyte
compared to regions away from it. The free energy of this phase
can be approximated by separating a region near the polymer from
the rest of the solvent. Since the polymer concentration is very
dilute, this region can be considered as a separate thermodynamic
system in equilibrium with an ion reservoir. The amount of charge
per monomer associated to the polymer from each of the ionic
species is $f_{m}$, $f_{u}$, $f_{t}$. The average ionic
concentrations in the bulk of the solution excluding regions
occupied by polymer is then $(m,t,u)_{eff}=(m-f_{m}p/z,\,
t-f_{t}p,\, u-f_{u}p)$.

The free energy density for the system polyelectrolyte plus
solvent can be written as:
\begin{equation} F_{t}=\frac{\phi}{N}\ln\frac{\phi}{N}+ \phi F_{e} + F_{s}, \end{equation}
where $F_t$ is the total free energy density in the dilute phase,
$F_e$ is the energy per electrolyte monomer due to the
interactions between its own and associated charges, and $F_s$ is
the energy of the solvent an the ions in solution. The values of
the associated fractions $f_{i}$ are determined by minimization
of this expression with respect to them.

The salty environment can be modeled as an ionic solution obeying
a Debye-Huckel free energy where it is necessary to consider not
only the multivalent positive, monovalent positive  and negative
ions but also an extra species, the reassociated ions. At the
typical concentration at which the redissolution of the
electrolyte takes place, there are important corrections to the
free energy of the solution due to this phenomenon. The simplest
way to describe the association is to assume that a finite
fraction $f_{b}$ of all the multivalent ions in solution are
associated with the same number of positive monovalent ions thus
forming a new species  of reduced charge $(z-1)$ with
concentration $m_a=f_{b}m_{eff}$. While in general there can be
further reassociation to states of charge $z-2$ etc., only the
first reassociation will be treated explicitly here. Upon
reassociation, the magnitude of the energy of a pair of associated
ions is $g_b$. The new concentrations of non associated ions are
$m_n=m_{eff}-m_a$, and  $u_n=u_{eff}-m_a$. The free energy for the
solvent is then:
\begin{equation}F_{s}=\sum\psi_{i}\ln\psi_{i}-\frac{1}{12\pi}\kappa^3- g_b m_a
\end{equation}
where  $\kappa$ is the screening length of the system and is given
by
\begin{equation}\kappa^2=\sum_{i}4\pi z_{i}^2\psi_{i},\end{equation}
where the ionic densities $\psi_i$ with charges $z_{i}$ are
$m_a$, $m_n$, $t_{eff}$ and $u_n$.

The free energy of the near region of the polyelectrolyte, per
monomer of electrolyte can be expressed as an expansion up to
quadratic order on the fractions of associated charges. A good
choice for this functional is
\begin{equation}F_{e}=\frac{G_2}{2} q^2%
-g_m f_m -g_t f_t -g_u f_u +g_cf_m f_u. \label{expansion}
\end{equation}
To argue for such functional it is convenient to separate within
the near region of the electrolyte, the contributions to the free
energy coming from a small region near each positive multivalent
or monovalent ion, and the interaction of between these ions with
others further away. The long range screened interaction between
charges gives rise to the quadratic contribution   $G_{2}q^2/2$,
where the effective charge per monomer is $q=1-f_m+f_u-f_t$. The
coefficient of interaction depends on the geometry of the system
and for simplicity is chosen to have the form $G_2=G_1
B\ln((1+\kappa)/\kappa)$, where $G_1$ is a geometric factor. For
large screening lengths $\kappa^{-1} \gg 1$, this term reflects
the interaction of each charge with the logarithmic field of a
charged rod of length $\kappa^{-1}$. At very small screening
lengths $\kappa^{-1} < 1$ the function goes to $0$.

When the effective electrolyte charge is near zero the linear
terms in the free energy reflect the correlation energy of each
charge with its near neighbors. At zero effective charge, the
amount of energy gained from the addition of a unit charge from a
given ion is simply $-g_m$, $-g_s$, or $-g_t$. The association of
a positive monovalent charge has as prerequisite the presence of
multivalent negative ions associated to the electrolyte, so that
$f_{u}<f_{m}$. The reduction in energy for this association is
required to be comparable (but smaller) to the energy of
association for these pairs of ions in the bulk so that
$g_{u}<g_{b}$ the positive charge is brought into contact with a
positive associated ion but it is also in the neighborhood of the
negatively charged polyelectrolyte.

The reassociation of the multivalent ions and negative coions
influences the strong correlations between associated multivalent
ions. As a larger number of monovalent ions enter the near region,
the multivalent ions reduce its correlations since their repulsion
is diminished by the presence of the negative ions associated with
them. This element is reflected in the correlation reduction term
$g_c$. More such terms describing corrections to the correlations
can be written but it turns out that the inclusion of just this
one is sufficient to recover a phase diagram that is similar to
those observed in experiments. Omission of this term leads to much
more extended regions of coexistence.

\section{Determination of Coexistence Regions}

\subsection{Equilibrium between phases}
A state with concentrations $(p,m,t)$ decomposes into two states
if the following free energy, for some value of $x$, $0\leq x \leq
1$, is smaller than the total energy of that state.
\begin{equation}%
F_{comp}=xF_d+F_t(\phi-x\Phi, \rho-x\Phi/z, s+\Phi)\leq
F_{t}(\phi,\rho,s).\label{cond} %
\end{equation}
The boundary of the coexistence region is determined by points at
which the condition above is an equality and the required value of
$x$ is zero. Expansion of $F_{comp}$ with respect to $x$ leads to
\begin{equation}
-M-\mu<0
\end{equation}
where the effective chemical potential pre monomer for the
addition of a neutral electrolyte complex into the dilute region
is
\begin{equation}%
\mu=\partial_\phi F_{t}+(1/z)
\partial_\rho F_{t}-\partial_{s}F_{t}. %
\end{equation}
On the basis of the assumption made, the final condition for the
determination of the coexistence boundary does not explicitly
depend on the concentrations in the dense phase given by $\Phi$,
but only on the bulk energy per electrolyte charge $-M$ and the
slope of the tie-line. The specific value of $\Phi$ is,
nevertheless, important to determine the condensed electrolyte
fraction $x$.

\subsection{Collapse}

In a typical experiment performed to study the phase transitions
of this type of system, the amount of multivalent salt present is
changed from $0$ to some maximum value in small increments. The
first transition occurs (for low monovalent salt) when the amount
of charge provided by the multivalent salt is approximately the
same as the amount of polyelectrolyte. If there is a large amount
of monovalent salt, the transition requires a larger amount of
multivalent salt. Thus, it is possible to approximate the location
of the transition as follows: the condition Eq.\ref{cond} is
satisfied almost as soon as full charge compensation of the
electrolyte by multivalent ions occur. The chemical potential
$\mu_{m}$ of the ions in the solution, disregarding the Debye
Huckel contribution, is $\ln(m-f_{m}\phi)$ with $f_{m}\approx 1$.
Thus, for low monovalent salt concentrations the condition is
simply that there must be enough ions to compensate the bare
charge of the electrolyte:
\begin{equation} z m\approx \phi.\end{equation}
At larger monovalent salt concentrations, the multivalent ions
must first displace associated positive monovalent ions from the
electrolyte. The reduction in the energy from the association of
the monovalent and multivalent ions is, per charge, approximately
the same and thus the chemical potential in the solvent for the
multivalent ions should be smaller than that of the monovalent
ions. The condition for the charging of the electrolyte with
multivalent ions, and a first approximation to the collapse
transition condition is:
 \begin{equation} \frac{1}{z}\ln m \approx \ln(s +\phi).\end{equation}

\subsection{Redissolution}

In solutions where even at large salt concentrations the free
energy of the ions is dominated by their translational entropy and
there is little correlation between charges, the redissolution
transition occurs when the decrease in energy per electrolyte due
to multivalent ions overcharge is larger than the bulk correlation
energy of the dense phase. However, if many monovalent ions are
present, the onset of association of the positive monovalent with
negative multivalent ions marks a sharp decrease in the energy of
the electrolyte.  Not only there is a decrease in the energy of
the positive charges but also it is possible to associate many
more ions. The repulsive interaction is decreased as the near
region can be near neutral while having acquired large number of
positive ions.

To first approximation, the onset of the association of positive
ions occurs when the magnitude of the translational entropy of the
free monovalent ions is comparable to the energy of association
$g_b$. At larger monovalent ion concentrations, the charging of
the electrolyte by positive multivalent ions together with
monovalent coions occurs when the energy of association $g_b$ is
capable of compensating the translational entropy of the
monovalent ions. Thus we have
\begin{equation} \ln u= \ln
(zm+s) \approx -g_{b}.%
\end{equation}
This expression provides a rough approximation for the location of
the second transition. For fixed amounts of added monovalent salt,
the transition location is approximately a horizontal line in the
($p$,$m$) or ($\phi$, $\rho$) plane.

\subsection{ A numerical example.}

The results of a numerical minimization of the model free energy
appear in Figure 3. There it is shown, for different values of the
monovalent salt concentration, the predicted regions of
coexistence in the $(\phi,\,\rho)$ plane. The parameters used for
this example are $B=3.0$, $g_m=9.0$, $g_u=9.0$, $g_t=9.6$,
$g_b=1.5$, $g_{c}=1.8$, $M=9.05$. The curve containing points
$a$, and $b$ has the lowest amount of added monovalent salt
$s=1.8\, 10^{-5}$ (intended to represent a 0.01 M molar
concentration), the other two curves are for concentrations
$s=9.0 \, 10^{-5}$, and $s=1.8 10^{-4}$ (0.05 M, and 0.1 M). The
largest salt concentration leads to smaller coexistence regions.
The marked pairs of  points $(a,b)$, $(c,d)$, and $(e,f)$
correspond to the location of the pairs of collapse-redissolution
transitions that occur when for fixed amounts of monovalent salts
and fixed amounts of polymer, multivalent salt is added to the
system.

\begin{figure}
\psfig{file=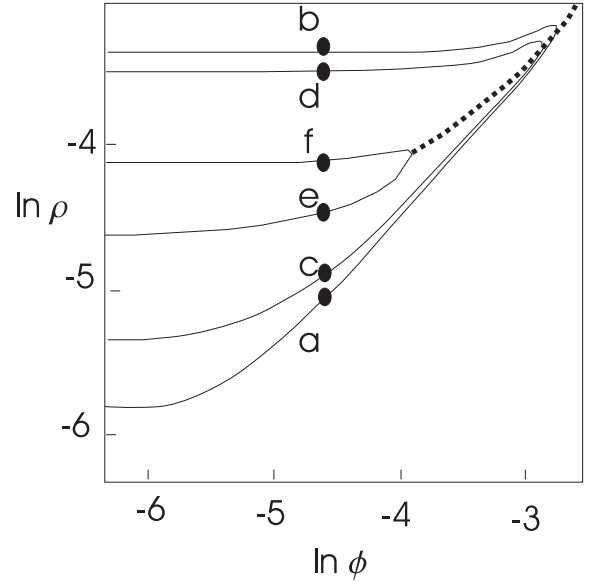,width=3.0in,angle=0} \caption{The three solid
curves correspond to the coexistence region boundary for three
sections of the phase diagram in the $(\phi, \rho, s)$
presentation, for different  amounts of monovalent salt, details
in text. The largest amount of salt produces the smallest
coexistence region. All states outside the boundary for each
section have a negative net charge. For the case with largest
amount of monovalent salt the dotted curve marks the border of the
region where the multivalent ions overcompensate the bare charge
of the electrolyte.}
\end{figure}

In this example, at the boundary of the redissolved region for the
different values of monovalent salt added, the states $d$, $e$ and
$f$ are all overcharged with positive multivalent ions but the net
charge of all these states is negative. For the choice of
parameters made, all states of the system that are not decomposed
(lying outside the coexistence curve for fixed amount of
monovalent salt) have a total negative charge. Different choices
of parameters can lead to different charged states and, for
example, in systems where the energy of reassociation of the
multivalent salt is very  small, all the dissociated states should
have a positive effective charge, and the phase diagram has the
shape proposed in work of Nguyen an Shklovskii \cite{nguyen},
where the lines of redissolution and collapse do not join to form
a bounded coexistence region. The model presented here also
predicts that there is a maximum value of monovalent salt for
which there is a coexistence region. As in the scheme of Figure 2,
the coexistence region has a limited extent in the $s$, $t$ or $u$
axes.

For the case of largest amount of monovalent salt added, (with
coexistence curve containing the points $(e,f)$), the dotted line
divides the region of states that, while effectively negative,
have enough multivalent ions to overcompensate the electrolyte
bare charge. Separating curves for other salt concentration show
similar behavior, always originating near the "tip" of the
diagram. The coexistence region sections are closed regions but
their "left" boundaries lie at very small concentrations that are
typically experimentally inaccessible, and are not shown in Figure
4.

 In extremely dilute solutions there is no large difference
between the original concentration of ions in the bulk of the
solution and the effective concentrations obtained by subtracting
the associated ions. As the amount of polymer increases, the
depletion from the bulk reduces the amount of available ions and
states with similar amounts of associated ions appear only at
higher multivalent salt concentrations. This leads to an upward
curving trend in the redissolution curves.

\section{Conclusions and Future work}

It is still an important but unfinished problem to refine the
description of the layer of associated ions by both providing a
concrete definition of the layer and by testing possible (mean
field) models of the free energy contribution of the layer to the
total energy of the system. In the context of integral equation
approaches this goal can be achieved by locating specific
properties of the ion distribution functions \cite{holm}, but
ideally such definition ought to be model independent and
experimentally useful. Nevertheless, even with arbitrary
definitions of the layer, a free energy expansion of the form  Eq.
\ref{expansion} is a useful instrument for the analysis of this
type of system.

One would like to reduce the large number of parameters used in
the present description of the system as many properties of real
systems cannot be precisely measured or have a poor experimental
definition. To use the ideas presented here in the context of the
real system requires the collapse of information about the
interaction between molecules and monomers into a few simple
constants. These quantities reflect the average radii of ions and
are supposed to include  the steric effects of water as well as
the specific location of charges in the electrolytes.  A careful
parametrization of simulation results, on the basis of the charge
of the near region of an electrolyte was, for example, presented
in the work of Messina {\it et al.} \cite{messina}. In model
systems studied though simulations or integral equations is
possible to reduce the number of these parameters and calculate
them on the basis of simpler geometric parameters. For comparison
with  experimental data it is, however, preferable to retain these
constants as adjustable parameters.

A couple experimental tasks are suggested by this work as
important to improve our understanding of these systems: the
presentation of the data available from the point of view of phase
separation of ionic species (instead of salts), a more systematic
mapping of the effective charge of the electrolytes in the
redissolved region of the phase diagram, and determination of the
ionic composition of the associated layer (in this respect there
has been progress by means of osmotic pressure measurements
\cite{amis,delsanti,claudine,raspaudosmotic}). Future analysis of
experimental data using the model presented in this article
should provide a map, even if only a rough one, of the properties
of specific ionic species with the parameters used. Finally, the
addition of further ingredients that can directly modify the
chemical potential of the different charged species should also
induce dramatic changes in the location a of the transitions.
From the point of view of the approach presented here, these
modifications would be reflected in changes of the
phenomenological parameters used in the model.

\section*{Acknowledgements}

The author thanks many useful talks with Monica Olvera de la Cruz,
Eric Raspaud, Boris Shklovskii, Toan Nguyen, Motohito Tanaka, Mark
Stevens, and John Widom.

\end{multicols}
\end{document}